EPILOGUE

G. Fabbiano

*Harvard-Smithsonian Center for Astrophysics, 60 Garden St., Cambridge MA 02138, USA*

**Abstract:** The IAU Symposium 230, Populations of High Energy X-ray Sources in Galaxies has been a wide-spectrum affair, with talks discussing results from the soft X-ray to the Gamma–ray range on virtually the entire universe, from our Galaxy to the high redshift regions when first galaxies emerged. I do not name any presenter in this summary, but concentrate on themes and results that I have found striking.

**1.0 From the Milky Way to the Deep Universe**

First, a few words on the present state of high energy studies of X-ray sources in galaxies, as it has unfolded during the symposium. In a way we have been doing a reverse peeling of the onion, starting from the core and moving outwards. I remember when *Uhuru*, the first X-ray satellite, produced the first map of the Milky Way in the 2-10 KeV range. Now similar maps are being produced with *Swift* and *Integral* in the hard X-rays/Gamma-ray range. Observations of the Local Group and nearby galaxies with *XMM-Newton* and at higher angular resolution with *Chandra* are generating and calibrating the tools for X-ray population studies in the X-ray range (~0.2-10 keV): X-ray photometry and Luminosity Function (XLFs). Comparison of these results with optical counterparts (either from detailed identification, or from positional association with the stellar population), are laying the foundations of X-ray population studies.

With the ability of surveying many galaxies provided by the current sensitive telescopes (*Chandra, XMM-Newton, Integral, Swift*), new classes of soft and hard X-ray sources have been studied and/or discovered. *Chandra*'s sub-arcsecond angular resolution and sensitivity allows for the first time a wide-ranging study of quiescent supermassive nuclear black holes and their environment, to probe the inefficient fueling stage of AGNs and dark outflows. *Integral* and *Swift* are exploring the association of Gamma-ray bursts with hypernovae. *Chandra* and *XMM-Newton* are probing the nature of ULXs and their debated association with intermediate-mass black holes; a widespread association of these luminous sources with star-forming galaxies was discovered with these X-ray observations and has raised the possibility of young normal X-ray binary counterparts.

This is the coming of age of X-ray population studies. We are moving from merely seeking to understand how individual engines work, to study the collective properties of different high-energy sources in different stellar populations. *Integral* is pioneering these studies in the Milky Way at the high energies, retracing the work done in the past 30 years at energies below 10 keVs. The high resolution of *Chandra* is needed to extend these studies to a large number of galaxies out to the Virgo cluster and beyond. These results are reopening some long shelved question on the formation and evolution of X-ray binaries, especially that of the role of globular clusters in LMXB formation, and fostering the beginning of X-ray population synthesis studies.

The next step is from X-ray source population studies to galaxy evolution. Again the resolution of *Chandra* is needed to go deep in the universe and study the X-ray evolution of galaxies with redshift. *Chandra* and future high resolution sensitive X-ray missions will be needed to pursue these studies in individual distant galaxies and probe the formation and evolution of intermediate-mass and supermassive black holes early in the history of the universe.

This symposium has presented us with two main unifying themes and a puzzle. The first theme is the quest for a unified scheme for all accretion sources, from X-ray binaries to quasars, including ULXs and quiescent galactic nuclei. We seek to understand the nature of the compact accreting object and relate the emission properties to the fuel supply. The second theme is the connection between high-energy sources and stellar evolution, including both the formation and evolution of X-ray binaries in different stellar populations and environments, and the link of Gamma-ray bursts with supernovae and hypernovae, that in some cases may be a step in the formation of compact accreting binaries.

The puzzle is: do we have Population III stars? Are these the progenitors of ULXs? Can we detect their remnants or constrain their numbers? How do X-ray observations compare with the theoretical expectations of black hole formation and merging evolution?

## 2. What's Next?

Deepening our understanding of these results will require future high-energy observations, both with the present complement of telescopes, but also with future missions. If we want our field to fulfill the promises of the present results, a new generation of observatories is needed. I want to point out that *Chandra* has been unique in opening up the study of the evolution of X-ray populations. These studies, both in nearby galaxies, and (collectively) in the deep universe, need the sub-arcsecond resolution of *Chandra*. I have long been concerned that the future X-ray missions being planned do not include a high resolution observatory to take the place of *Chandra* were it to stop operating. Many colleagues, who are 'dependent' on *Chandra*, have expressed this sentiment (without my prompting, honest!) during coffee breaks. I urge that a *Chandra-2* study be considered, in parallel with the planned *Constellation-X* (NASA) and *XEUS* (ESA) missions.

*Constellation-X* and *XEUS* will explore the spectral and timing domains, building up on the *XMM-Newton* work, but with significantly larger collecting areas. These missions will certainly improve our understanding of physics in extreme circumstances, but are not designed to follow up the high resolution work begun with *Chandra*. The *Generation-X* concept, presently in study by NASA, would unite a very large (~100 square meters) collecting area with an angular resolution comparable with that of the *Hubble Space Telescope*, and would be the ultimate machine for X-ray population studies, but this is way in the future.

Meanwhile, our community must preserve and find better way to exploit the available data. The high energy community has a tradition in this sense, recognizing the uniqueness of the space observations and the effort required to collect these data. This archival tradition fits well with the world-wide 'Virtual Observatory' movement. Under the umbrella of the International Virtual Observatory Alliance (IVOA), which coordinates national efforts (e.g., the National Virtual Observatory – NVO, in the US), the astronomical community is seeking to facilitate the access to multi-wavelenght data archives and data analysis tools. The high energy community is integral part of this effort and is leading in the establishment of astrophysical data models, which provide the foundation for these developments and for future mission-independent data analysis tools.

To conclude these parting reflections, I would like to thank all the people who have made this Dublin IAU Symposium 230 possible and highly successful. This is a long list, including the Local Organizing Committee, and staff and students of University College Dublin and Dunsink Observatory. Their hard work and hospitality has made this fruitful meeting possible. I would also like to thank the Irish Government and the Irish Ministry of Science for their hospitality and support of this meeting. Finally, I thank the CXC for providing personal travel support under NASA contract NAS8-39073.